\documentstyle[twocolumn]{article}
\newlength{\aivwidth}   
\newlength{\tmpwidth}   

\title{ Vortices in superconducting films are anyons }
\author{Jacek Dziarmaga  \\
        Jagellonian University, Institute of Physics, \\
        Reymonta 4, 30-059 Krak\'ow, Poland
        \thanks{e-mail address: ufjacekd@ztc386a.if.uj.edu.pl}
        \thanks{address from October 1, 1995 -
                Department of Mathematical Sciences,
                University of Durham, Durham, DH1 3LE, United Kingdom}}
\date{April 20, 1995}
\begin{document}
\maketitle
   \begin{abstract}
  We point out that nontrivial quantumstatistics of vortices
in planar superfluid systems has its origin in the Schrodinger
equation rather then the Chern-Simons term. Vortices in superfluid
helium films are not anyons because helium films are compressible.
Quantum Hall effect layers admit vortices with nontrivial statistical
interactions thanks to their incompressibility. We consider type
II planar superconductors described by time-dependent nonlinear
Schrodinger equation. Inside of a vortex core superconductor
is in the normal state. Because of that vortices in superconducting
films are anyons. We consider also vortex loops in 3 dimensions
but a phase factor due to linking of vortex loops' worldsheets
is not a topological invariant.
   \end{abstract}

\vspace*{0.5cm}
TPJU 9/95\\
cond-mat/9504???\\
\vspace*{0.5cm}
\newcommand{\be}{\begin{equation}\label}
\newcommand{\ee}{\end{equation}}
\newcommand{\ba}{\begin{eqnarray}\label}
\newcommand{\ea}{\end{eqnarray}}
\newcommand{\pl}{\partial}
\newcommand{\lb}{\lambda}
\newcommand{\kp}{\kappa}
\newcommand{\tr}{\triangle}
\newcommand{\dt}{\delta}
\newcommand{\bt}{\beta}
\newcommand{\al}{\alpha}
\newcommand{\gm}{\gamma}

  Interactions of vortices in superfluids described
by the nonlinear Schrodinger equation are by now a well studied problem
\cite{neu,lund,lee,top,close}. What has not been studied in detail is the
quantum theory of vortices. In \cite{top} there has been proposed a topological
Landau-Ginzburg theory for vortices in superfluid helium. The theory
is equivalent to the standard nonlinear Schrodinger equation provided
the topology of vortex lines is determined. The authors of \cite{top} have
derived, in 3-dimensional case, Berry phase picked up by the microscopic
wave-function due to vortex-loops' motion.

  In this paper we consider mainly planar systems which may be
relevant to description of thin superfluid films. It is shown that
there is a Berry phase proportial to the area enclosed by the vortex
trajectory. Mutual interactions are considered and it appears
that vortices might be anyons if a superfluid helium film were not
compressible.
Thus we try to find a model with an incompressible condensate.
The phenomenological models for quantum Hall effect appear to be a well
known example. However we argue that Chern-Simons term is not essential
for transmutation of vortices' quantumstatistics. In fact the statistical
interactions come solely from the Schrodinger term. Finally we
consider type II superconductors. Thanks to the normal current
carriers inside a vortex core a defect of superfluid charge
inside a vortex with respect to uniform background can be nonzero.
It is shown that vortices in superconducting films are anyons.

\paragraph{ Vortices in superfluid helium films. }

In the local limit the Lagrangian density of the model for interacting
hard-core bosons is \cite{pit}
\be{100}
L=i\hbar\Psi^{\star}\pl_{\tilde{t}}\Psi
-\frac{\hbar^{2}}{2m}\tilde{\nabla}\Psi^{\star}\tilde{\nabla}\Psi
 +\mu\Psi^{\star}\Psi-\frac{1}{2}\lb(\Psi^{\star}\Psi)^{2} \;\;,
\ee
where $\Psi$ is the condensate wave-function defined on the plane.
We assume quantisation of the model by path integrals does make sense.
It is convenient to rescale the fields and coordinates in the above
formula
\be{110}
\Psi=\sqrt{\frac{\mu}{\lb}}\psi \;\;,\;\;
\tilde{t}=\frac{\hbar}{m} t \;\;,\;\;
\vec{\tilde{x}}=\frac{\hbar}{\sqrt{m\mu}}\vec{x} \;\;
\ee
and subsequently multiply the whole Lagrangian density by
$\frac{m\lb}{\mu\hbar^{2}}$ to obtain the dimensionless Lagrangian
\be{120}
L=i\psi^{\star}\pl_{t}\psi-\frac{1}{2}\nabla\psi^{\star}\nabla\psi
 -\frac{1}{2}(1-\psi^{\star}\psi)^{2}                                \;\;.
\ee
A field equation is
\be{130}
i\pl_{t}\psi=-\nabla^{2}\psi+(\psi^{\star}\psi-1)\psi \;\;.
\ee
The model admits vortex solutions of the form
$\psi=h_{n}(r)\exp in\theta$. $n$ is the winding number. The profile
function satifies the equation
\be{140}
h''_{n}+\frac{h'_{n}}{r}-\frac{n^{2}}{r^{2}}h_{n}+(1-h^{2}_{n})h_{n}=0 \;\;.
\ee
Asymptotices close to the origin and at infinity are respectively
\be{150}
h_{n}(r)\sim h_{n}^{0}r^{n}+O(r^{n+2}) \;\;,\;\;
h_{n}(r)\sim 1-\frac{n^{2}}{2r^{2}}+O(r^{-4}) \;\;.
\ee
Energy of the vortex is regular at the origin but logarithmically
divergent at infinity. The divergence can be regularised by putting
the system into a finite-size box. Energy of a vortex-antivortex
pair is also finite since the topology of such a configuration
is trivial - there is no slowly falling down gradient of the phase of
the scalar field at infinity.

\paragraph{ Effective Lagrangian for vortices. }

Let me first describe derivation of an effective Lagrangian for
widely separated vortices. The field configuration can
be well approximated by the product Ansatz
\be{210}
\psi=\prod_{v}h[\vec{x}-\vec{X}_{v}(t)]
     \exp[i\al_{v}\Theta(\vec{x}-\vec{X}_{v}(t))] \;\;,
\ee
where the product runs over particular vortices. $\al_{v}$ is a sign
of circulation of the $v$-th vortex and $\theta(\vec{x})$ is the
azimutal angle around zero of its argument.

  The Lagrangian density (\ref{120}) can be rewritten as
\be{220}
L=-\rho\pl_{t}\chi-\rho(\nabla\chi)^{2}-(\nabla\rho^{\frac{1}{2}})^{2}
  -\frac{1}{2}(1-\rho)^{2} \;\;,
\ee
where we have replaced $\psi=\rho^{\frac{1}{2}}\exp i\chi$. With the use
of the product Ansatz (\ref{210}) the Lagrangian takes the form
\be{230}
L=-E_{0}-\int d^{2}x[\rho\pl_{t}\chi
  +\rho\sum_{v\neq w}[\al_{v}\al_{w}\nabla\Theta(\vec{x}-\vec{x}_{v})
                      \nabla\Theta(\vec{x}-\vec{x}_{w})]] \;\;.
\ee
The first term is minus the net energy due to nonzero winding number
and to core contributions.
The second term contains information about vortex interactions.
This term is regularised by $\rho$ - the modulus of the scalar field.
The modulus is close to 1 almost everywhere except the cores of vortices.
Loosely speaking the phase gradients and the time derivative of the phase
in the second term have to be integrated over the whole plane except
vortex cores. The first term to be integrated can be rewritten as
\be{240}
-\sum_{v}\al_{v}\int d^{2}x\;\rho\pl_{t}\Theta[\vec{x}-\vec{x}_{v}(t)] \;\;.
\ee
If there were only one vortex a careful treatement of the integration of the
multivalued phase would lead to \cite{lund}
\be{280}
L^{eff}_{1}=-\pi\al_{v}\vec{X}_{v}(t)\times\dot{\vec{X}_{v}}(t)
\ee
Upon quantisation by path integrals this factor would lead to a phase
picked up by the wave-function. If a vortex trajectory
were closed than
\be{290}
S^{eff}_{1}=\int_{t_{1}}^{t_{2}}L^{eff}_{1}=
             2\pi\al_{v}\times\;"area\;enclosed\;by\;trajectory" \;\;.
\ee
The area should be taken with a positive sign for a clockwise motion
and with a negative sign for an anticlockwise motion. For trajectories
like, say, the numeral "8" one part of the contour gives a positive
contribution and the other one negative. An
equivalent interpretation is that the phase picked up by the wave function
is proportional to the total mass of the superfluid enclosed by the contour.
This is the whole story in the one-vortex case.

  If there were more than one vortex an additional term would arise
from (\ref{240}) in addition to (\ref{290})
\be{300}
S^{eff}_{2}=M_{0}\int_{t_{1}}^{t_{2}}
    \sum_{v>w}\al_{v}\al_{w}
                \frac{d}{dt}\Theta[\vec{x}_{v}(t)-\vec{x}_{w}(t)] \;\;,
\ee
where
\be{310}
M_{0}=2\pi\int rdr\;[1-h_{1}^{2}(r)]  \;\;
\ee
is the defect of the superfluid mass with respect to the uniform
background. With the asymptotics (\ref{150}) it becomes clear that
the integral is logarythmically divergent. Thus there is no dilute
vortex gas limit in superfluid helium in which vortices might be anyons.

\paragraph{ Discussion. }

   There is no long-range statistical interaction between vortices
in the nonlinear Schrodinger equation. Compressibility of
superfluid manifests itself in divergence of the integral in Eq.(\ref{310}).
In the incompressible case this integral would be convergent but,
as it was shown in \cite{hawu}, two-dimensional superfluid films are
compressible. It is so because of the contribution to the
energy density from the term $\mid\nabla\chi\mid^{2}$ which is nontrivial
in the case of nonzero net superfluidity. In other words, the term
$\frac{n^{2}}{r^{2}}h_{n}$ in Eq.(\ref{140}) gives rise to the long-range
term in the second of Eqs (\ref{150}). This term might be matched if the local
interaction in the model (\ref{100}) were replaced by a two-body
repulsive potential $V(\vec{R})=\frac{g}{\vec{R}^{2}}$ with some critical
value of the constant $g$. Such a model would have nothing to do
with superfluid helium as long-range interactions are attractive there.
However it shows that one can have anyons in a model just because
of the Schrodinger term and without any explicit parity-breaking
terms like Chern-Simons interaction. The Schrodinger term
has a remarkable form very similar to the Berry phase in Quantum Mechanics.
If we consider evolution of the condensate wave-function
due to slow motion of a vortex the contribution of the Schrodinger
term to the effective action is
\be{590}
i\int_{t_{1}}^{t_{2}}dt\; \dot{\vec{X}}
<\psi(\vec{x},\vec{X}),\frac{d}{d\vec{X}}\psi(\vec{x},\vec{X})> \;\;,
\ee
where $\psi$ is the condensate wave-function dependent
on a vortex position $\vec{X}$. $<>$ is the standard scalar product.

  Thus an obstacle is identified. The core must be well-defined and
for the core to be well-defined we have to make the fluid less compressible.
The contribution from the phase-gradient energy can be matched by
gauging the model $\pl_{\mu}\chi\rightarrow\pl_{\mu}\chi+iA_{\mu}$.
An example are the models with Chern-Simons interaction being
a phenomenological description of the quantum Hall effect. In these
nonrelativistic models \cite{jackiwpi,ezawa,igor,ja3} the scalar field
of the nonlinear Schrodinger equation is minimally coupled to the C-S field
\be{600}
\pl_{\mu}\psi\rightarrow \pl_{\mu}\psi-ia_{\mu}\psi \;\;.
\ee
The kinetics of the gauge field is governed by the Chern-Simons term
\be{610}
\kp\varepsilon^{\mu\nu\al}a_{\mu}\pl_{\nu}a_{\al}  \;\;.
\ee
All the cited models possess the Bogomolny limit \cite{bogomolny} and
in this limit there are static multivortex solutions.

   The dynamics of CS vortices in the Bogomolny limit
has been intensively studied in relativistic models \cite{kimmin,cs}
and also in the nonrelativistic context \cite{liu,ja3}. One starts from
a static multivortex solution in the Coulomb gauge. In this gauge
the phase of the scalar field is a sum of azimutal angles exactly the same
as in (\ref{210}). Since the gauge field is transverse the CS term
does not contribute to the effective Lagrangian. The only contribution
comes from the Schrodinger term $i\psi^{\star}\pl_{t}\psi$. This is exactly
the same as in the nonlinear Schrodinger equation. A different choice
of the gauge would not change the topology of the phase of the scalar
field - the nontrivial phase factor for a closed trajectory would still come
solely from the Schrodinger term. Thus the nontrivial statistics
of vortices comes from the Schrodinger term and not from the CS term.
The CS term transmutes the quantumstatistics of the elementary field
quanta only. The statistical interaction between vortices arises
because there is a definite mass defect inside the closed trajectory
of a vortex (or a mass excess for Jackiw-Pi solitons).
The Gauss' law establishes connection between matter density
and magnetic field
\be{620}
\kp B=\psi^{\star}\psi \;\;.
\ee
The mass defect is quantised because the magnetic flux of the vortex
is quantised. The interaction term at long distances is up to a numerical
factor proportional to
\be{630}
\kp \sum_{v>w}\al_{v}\al_{w}
              \frac{d}{dt}\Theta[\vec{x}_{v}(t)-\vec{x}_{w}(t)] \;\;.
\ee
The prefactor $\kappa$ is the same as in the CS term (\ref{610}).
This coincidence might suggest that the CS term is responsible
for the statistical interaction between vortices. In fact it is rather
a dynamical effect of the CS term - in the limit of very small $\kp$
vortices become very thin - the mass defect goes to zero. On the other
hand for $\kp=0$, in the nonlinear Schrodinger equation, the core
width is infinite. The limit $\kp\rightarrow 0$ is singular - one
can not analytically continue Eq.(\ref{630}) to vanishing $\kappa$.
The crucial effect of the Chern-Simons gauge field is to make
the mass defect finite - all the fields around a vortex are well
localised.

   Vortices appear to be anyons in the gauged nonlinear Schrodinger
equation with Chern-Simons or Chern-Simons-Maxwell term. One might
wonder if it is also possible
to have anyons with just the Maxwell term. The answer is negative.
The phase factor is proportional the mass missing in the vortex core
($\rho$ is less than 1 there). However in the Maxwell-Schrodinger case
$\rho$ is also charge density and because of screening
the net charge (mass defect) of a vortex must be zero. Addition
of the Chern-Simons term modifies the Gauss' law and nonzero net charge
becomes possible.

\paragraph{ Vortices in type II superconductors. }

   The fact that Maxwell-Schrodinger dynamics does not admit anyons
does not rule out anyons in real superconductors. The evolution
of the condensate wavefunction is described by such a model
but there are normal charges and currents in addition to those of the
condensate.
Because of screening the total net vortex charge has to vanish.
Such a weakened restriction does not rule out the possibility of nonzero net
superconducting charge and that is what really matters. Below we will
derive statistical interaction between vortices with an assumption of local
charge
neutrality but this assumption can be released in the dilute vortex
gas limit.

 The time-dependent Ginzburg-Landau model is defined by the following
Lagrangian density
\ba{sup10}
L&=&-\frac{1}{4}F_{\mu\nu}F^{\mu\nu}
 +\frac{1}{2}i(\psi^{\star}D_{0}\psi-c.c.) \nonumber\\
 &-&\frac{1}{2}D_{k}\psi^{\star}D_{k}\psi
 -\frac{1}{8}\lb(1-\psi^{\star}\psi)^{2}
 +A_{0}\rho_{b}+A_{\mu}J^{\mu}_{n} \;\;.
\ea
The Lagrangian is already in rescaled dimensionless units. We will
consider this model as $2+1$ dimensional and later on discuss
modifications in for real thin superconducting films. $\lb$ is
a dimensionless constant which in the considered case of type II
superconductors is larger then 1. The covariant derivatives
are $D_{\mu}\psi=\pl_{\mu}\psi+iA_{\mu}\psi$. $\rho_{0}$ is
an uniform positive background charge density which in our rescaled
model is equal to $+1$. $J_{n}$ is the mentioned normal current
minimally coupled to the gauge field.

  Variation with respect to $A_{0}$ leads to Gauss' law
\be{sup20}
-\nabla^{2}A^{0}+\pl_{t}(div\vec{A})=\rho_{b}-\psi^{\star}\psi+J^{0}_{n} \;\;.
\ee
Now we assume local charge neutrality. In other words we assume
the distribution of the normal current is governed by some fast
degrees of freedom and it quckly adjusts itself so that the R.H.S.
of the above equation vanishes locally. This assumption is far too strong
for our needs and it is going to be released.
With this assumption we can choose the Coulomb gauge
$div\vec{A}=0$ together with $A_{0}=0$. We can forget now about
the scalar potential and thus the model in a static case reduces
to the Ginzburg-Landau theory
\be{sup30}
-L=\frac{1}{2}F_{12}F_{12}+\frac{1}{2}D_{k}\psi^{\star}D_{k}\psi
 +\frac{1}{2}\lb(1-\psi^{\star}\psi)^{2} \;\;.
\ee
It is well known \cite{ano} that such a model admits well localised vortex
solutions. Field equations are
\ba{sup40}
\nabla^{2}A^{l}=\frac{1}{2}i(\psi^{\star}D_{l}\psi-c.c.) \;\;,\nonumber\\
D_{k}D_{k}\psi=\frac{1}{2}\lb(\psi^{\star}\psi-1)\psi \;\;.
\ea
With an Ansatz
\be{sup50}
\psi(r,\theta)=h_{n}(r)e^{in\theta} \;\;,\;\;
A^{\theta}(r,\theta)=\frac{n}{r}A_{n}(r) \;\;
\ee
a vortex solution with a winding number $n$ is obtained. Both the scalar
field and the magnetic field are exponentially localised. For $\lb>1$
the penetration lenght of the magnetic field is larger then the coherence
lenght of the scalar field. The gauge invariant combination
$\pl_{k}\psi+iA_{k}\psi$ is exponentially falling down with a distance
from a vortex core. Multivortex solutions can be approximated
by a product Ansatz the same as in Eq.(\ref{210}) supplemented
by an additive Ansatz for the gauge fields.

  Once again we can consider Berry phases picked up by the wave-function
thanks to the Schrodinger term. This term is still the only term
which can contribute to the Berry phase because the only other
term with time derivatives in the Lagrangian (\ref{sup10}) is already
quadratic: $\frac{1}{2}\pl_{t}A_{k}\pl_{t}A_{k}$. Thus the considerations
as to the Berry phase in the second section can be directly applied to the
present model with a result
\be{sup60}
S^{eff}_{2}=M_{0}\int_{t_{1}}^{t_{2}}
\sum_{v>w}\al_{v}\al_{w}\frac{d}{dt}\Theta[\vec{x}_{v}(t)-\vec{x}_{w}(t)] \;\;,
\ee
where this time $M_{0}$ is a well-defined integral
\be{sup70}
M_{0}=2\pi\int_{0}^{\infty}rdr\;[1-h^{2}_{1}(r)]\equiv\pi R^{2}_{c} \;\;,
\ee
being a total mass defect of the superfluid directly related to the
missing charge $Q_{0}=-M_{0}$. The last equivalence in Eq.(\ref{26})
is a definition
of the radius of the vortex core. In this model vortex is a composite
of a nonzero net superfluid charge and a nonzero net magnetic flux.
One might think that this is the origin of the Aharonov-Bohm
effect. In fact it is just a mere coincidence.

Another effect considered before in a microscopic setting \cite{ao} is
a Berry phase for a single vortex (\ref{280},\ref{290})
\be{sup80}
S^{eff}_{1}=\int_{t_{1}}^{t_{2}}L^{eff}_{1}=
             2\pi\al_{v}\times\;"area\;enclosed\;by\;the\;trajectory" \;\;.
\ee
This Berry phase is responsible for the Magnus force.

These are not all the terms in the effective action. Another term is
a potential interaction derived in \cite{peru} which is exponentially
suppressed for widely separated vortices.
Finally, as we already have mentioned, there are terms in the Lagrangian
(\ref{sup10}) quadratic in time derivatives
$\frac{1}{2}\pl_{t}A_{k}\pl_{t}A_{k}$. These terms contribute to a term
in the effective Lagrangian quadratic in velocities
\be{sup100}
S^{eff}_{4}=\int_{t_{1}}^{t_{2}}dt\;\sum_{v}\frac{1}{2}m_{eff}
            \dot{\vec{X}}_{v}\dot{\vec{X}}_{v} \;\;.
\ee
One could naively think that $m_{eff}$ can be found by substituting to
$\frac{1}{2}\pl_{t}A_{k}\pl_{t}A_{k}$ multivortex fields with
time-dependent positions of vortices. In fact one would have to take
into account deformations of the multivortex fields for a given
trajectory up to terms linear in velocities. The calculations are involved
but they have been performed for relativistic Chern-Simons vortices \cite{cs}.
We are going to persue this topic in a separate publication. There is
an intriguing possibility that $m_{eff}$ can be determined by the
parameters of the model (\ref{sup10}).

  Thus we have shown that unlike vortices in superfluid helium films
those in superconducting films are anyons. In fact we have considered
a strictly $2+1$ dimensional model. In reality superconducting
films are immersed in 3-space. Vortices in such quasi-planar systems
are not so well localised. The combination $\pl_{k}\psi+iA_{k}\psi$
outside of the core does not fall down exponetially but rather
like $\frac{1}{r^{2}}$ \cite{degennes}. In superfluid helium $\pl_{k}\psi$
falls like $\frac{1}{r}$. The decay rate in a superconding film is strong
enough for the integral in Eq.(\ref{sup70}) to remain convergent. Thus our
conclusions are qualitatively unchanged for thin superconducting
films.

  Finally let me stress that the local charge neutrality assumption
is not essential. The essential point is that there is a nonzero
net superconducting charge associated with a vortex. This net charge
is enclosed by the trajectory of another vortex. It does not matter
if there are any local charge fluctuations provided they are localised
close to the core of the vortex.

\paragraph{ Extension to 3+1 dimensions. }

     So far we have considered only planar systems. Our goal was
to make our discussion somewhat parallel to the literature around
the quantum Hall effect. However because our anyons arise without
any help from any Chern-Simons term one is free to think about
generalisations to 3+1 dimensions. Of course there is a trivial
embedding of planar solutions in 3-space. There should be statistical
interaction between parallel vortex lines.
{}From a theoretical point of view statistical interaction between
vortex loops would be more interesting. One could consider closed
worldsheets of the loops. If it were not relevant one could similarly
as on a plane define a reference loop and close any open
worldsheet by continuing its boundaries to the chosen reference loop.

     Similarly as in the planar case one can work out the Berry phase
\cite{top} for vortex loops parametrised by time $t$ and lenght-like
parameters $\sigma_{l}$
\be{910}
S^{eff}=\frac{2\pi}{3}\sum_{l}\int dtd\sigma_{l}\;
        \vec{X}_{l}(\dot{\vec{X}_{l}}\times\vec{X}_{l}') \;\;,
\ee
where $l$ runs over the loops. " ' " means differentiation with
respect to the appropriate parameter $\sigma$. The parametrisation by
$\sigma$'s is so chosen that the circulation of the phase is clockwise
when one looks along the tangent to the loop.
This is the form of the Berry phase in the limit of vanishing core
thickness. The modulus of the phase
for a closed loop worldsheet is $2\pi$ times the volume enclosed by the
worldsheet. Similarly as on the plane one has to be careful with its sign.

  Mutual statistical interactions arise when one takes into account
the finite core thickness. The general formula for the Berry phase
in a compressible case (finite core thickness) reads \cite{top}
\be{920}
\frac{1}{2}\sum_{l}\int dtd^{3}x\;\int d\vec{X}_{l}
(\dot{\vec{X}}_{l}\times\frac{\vec{x}-\vec{X}_{l}}
{\mid\vec{x}-\vec{X}_{l}\mid^{3}})\rho(t,\vec{x})\;\;.
\ee
If the core thickness were neglected one could put $\rho=1$ everywhere
to obtain Eq.(\ref{910}).
Thus the modification to Eq.(\ref{910}) comes only from vortex cores
where $\rho$ deviates from the vacuum value. The Berry phase in 3
dimensions is not a topological index as can be seen in the following
example.

  Let us consider a pair of loops. One of them is a circle of radius R in the
x-y plane centered at the origin
\be{1010}
\vec{X}_{1}(t,\sigma_{1})=(R\cos\sigma_{1},R\sin\sigma_{1},0) \;\;.
\ee
The second loop is also a circle in a plane parallel to the x-y plane
but its radius and center are time-dependent
\be{1020}
\vec{X}_{2}(t,\sigma_{2})=[(R+a\cos\omega t)\cos\sigma_{2},
                           (R+a\cos\omega t)\sin\sigma_{2},
                           a\sin\omega t]\;\;.
\ee
The worldsheet of the second loop is a torus with the first loop inside
of it. The only contribution to the Berry phase comes from the motion
of the second loop and is equal to
$-2\pi\times ("volume\;of\;the\;torus"-2\pi^{2}RR_{c}^{2})$.
The second term is the volume of the thin vortex core of the first loop.
This term is an analogue of the statistical interaction of vortices
on a plane - it is the mass missing inside the closed
worldsheet. This term is not a topological invariant because
its value depends continuously on the radius $R$ while the topology
is $R$-independent.

\paragraph{Aknowledgement.}

This research was supported in part by the KBN grant No. 2 P03B 085 08
and in part by Foundation for Polish Science scholarship. I would like
to thank P.Ao for drawing reference [7] to my attention.


\end{document}